\documentclass[aps,prl,twocolumn,showpacs,groupedaddress,amsmath,amssymb]{revtex4}
\usepackage{graphicx}
\usepackage{color}
\usepackage{bm}
\usepackage{latexsym}

\begin{document}

\title{Light Localization induced by Random Imaginary Permittivities}
\author{A. Basiri$^{1}$, Y. Bromberg$^2$, A. Yamilov$^3$, H. Cao$^2$, T. Kottos$^{1}$}
\affiliation{$^1$Department of Physics, Wesleyan University, Middletown, CT-06459, USA}
\affiliation{$^2$Department of Applied Physics, Yale University, New Haven CT-06520, USA}
\affiliation{$^3$Department of Physics, Missouri University of Science and Technology, Rolla, MO-65409, USA}
\date{\today }

\begin{abstract}
We show the emergence of light localization in arrays of coupled optical waveguides with randomness only in the imaginary part of 
their permittivity and develop a one-parameter scaling theory for the normalized participation number of the Floquet-Bloch modes. 
This localization introduces a new length scale in the decay of the autocorrelation function of a paraxial beam propagation. Our results 
are relevant to a vast family of systems with randomness in the dissipative part of their impedance spatial profile.

\end{abstract}
\pacs{42.25.-p, 42.60.Da, 42.25.Bs }
\maketitle

Wave propagation in random media is of great fundamental and applied interests. It covers areas ranging 
from quantum physics and electromagnetic wave propagation to acoustics and atomic-matter wave systems. Despite this diversity, the underlying wave character of these systems provides a unified framework for studying mesoscopic transport and, in many occasions, points to new research directions and applications. 
A celebrated example of this universal behavior of wave propagation is the so-called Anderson localization phenomenon associated
with a halt of transport in a random medium due to interference effects originating from multiple scattering events \cite{A58}.  In recent years a number of experiments with classical \cite{WBLR97,CSG00,SGAM06,BZKKS09,HSPST08,C99,
LAPSMCS08,PPKSBNTL04,SBFS07} and matter waves {\cite{A08,I08} have confirmed the validity of this prediction. In all these 
cases, however, the wave localization originates from randomness pertaining the spatial profile of the reactive part of the impedance. 

\begin{figure}[tbp]
\begin{center}
\includegraphics[scale=0.325]{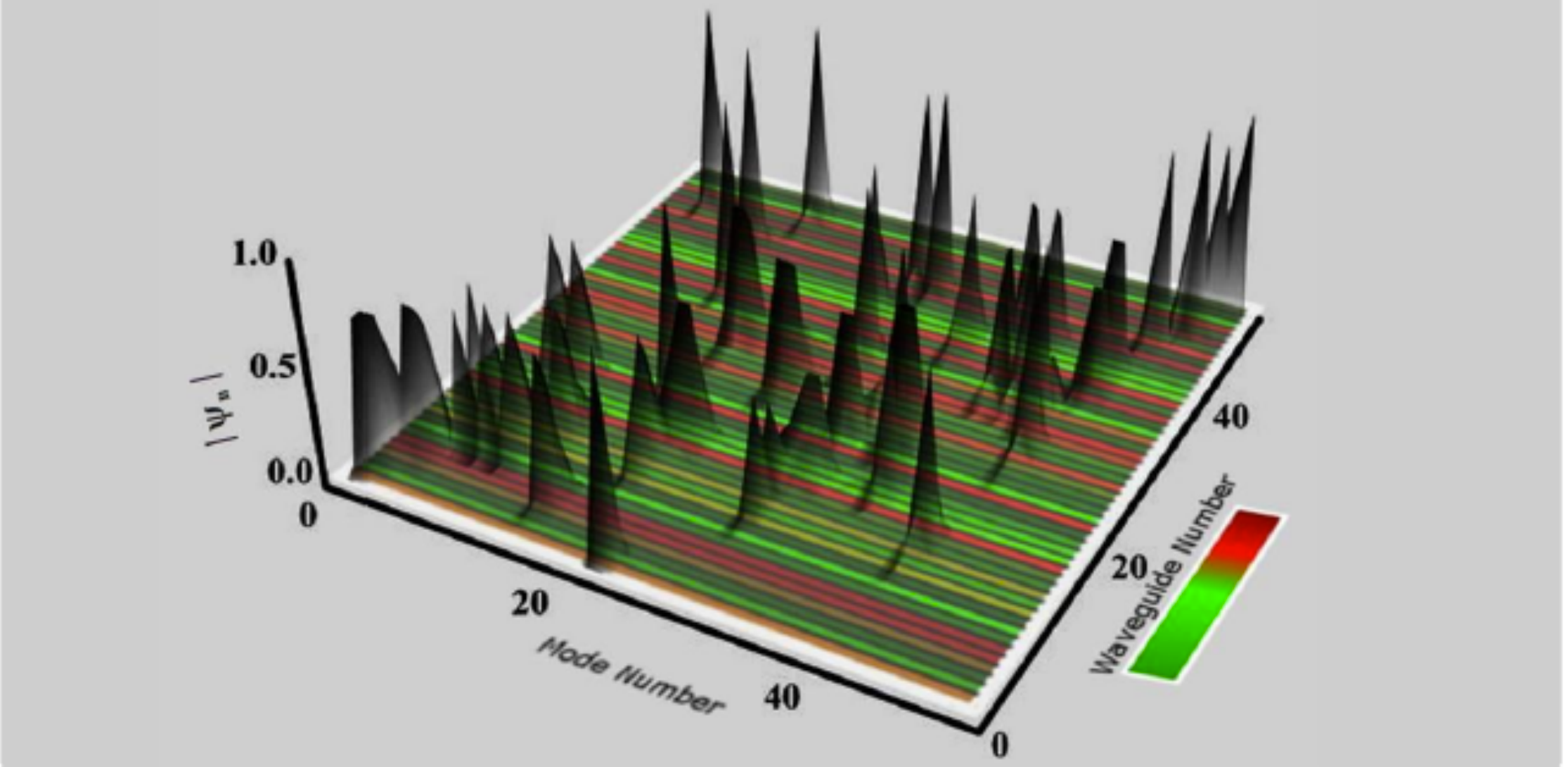}
\end{center}
\caption{The Floquet-Bloch modes of an array of $N=50$ waveguides with random imaginary index of refraction taken from a box distribution
with width $W=5$. All modes are exponentially localized at various localization centers corresponding to gain (red stripes) or loss (green stripes) 
waveguides alike.
}
\label{fig1}
\end{figure}

In the present paper we show the emergence of localization phenomena in a new setting, namely a class of systems, whose the spatial 
impedance profile has random dissipative part. Realizations of this class includes Bose-Einstein condensates in randomly leaking optical 
lattices, acoustic or electromagnetic wave propagation in a medium with random losses, and even quantum random walk protocols in 
the presence of traps that are used in the context of quantum computation. 

For concreteness we will refer below to a representative example of this class of systems drawn from optics: an array of $N$ 
coupled waveguides with complex index of refraction $\epsilon_n=\epsilon_n^{(R)}+i\epsilon_n^{(I)}$ where the real part $\epsilon_n^{(R)}$ 
can be the same for all waveguides while their imaginary part $\epsilon_n^{(I)}$ is a random independent variable given by some distribution. 
We find that the Floquet-Bloch (FB) modes $\Phi^{(\omega)}=(\phi_1^{(\omega)},\cdots,\phi_n^{(\omega)},\cdots)^T$ ($\phi_n^{(\omega)}$ 
is the amplitude of the FB mode at waveguide $n=1,\cdots,N$ associated with an eigenfrequency $\omega$) are exponentially localized 
with localization centers being waveguides with positive or negative imaginary refraction index alike. Specifically, we show that the averaged 
(rescaled) participation number $\xi_N(W,\omega)\equiv (\sum_{n=1}^N|\phi_n|^2)^2/\sum_{n=1}^N |\phi_n|^4$ obeys a one-parameter scaling:
\begin{equation}
{\partial p_N(W,\omega)\over \partial \ln N} = \beta \left(p_N(W,\omega)\right); \quad p_N(W,\omega)\equiv
{\langle \xi_N(W,\omega)\rangle \over N}
\label{eq1}
\end{equation}
Above $\beta$ is a {\it universal} function of $p_N(W,\omega)$ alone, 
and $\langle \cdots\rangle$ indicates an averaging over FB modes within a small frequency window and over disorder realizations. The 
variable $W$ defines the disorder strength associated with $\epsilon_n^{(I)}$ and it introduces a new length scale $\xi_{\infty}\equiv \lim_{N
\rightarrow \infty} \xi_N$ which is inversely proportional to the asymptotic decay rate of the FB modes. The transverse localization 
of the FB modes plays an important role in the beam dynamics. Specifically we find that the normalized autocorrelation function 
$C(z)\equiv (1/z)\int_0^z \left(|\psi_0(z')|^2 dz'\right)/\sum_n|\psi_n(z')|^2$ of a propagating beam $\psi_n(z)$ which is initially localized 
at waveguide $n_0$ deviates from its periodic lattice analogue at propagation distances $z^*\sim \sqrt{\xi/\Delta[{\cal I}m(\omega)}]$ 
where $\Delta[{\cal I}m(\omega)]$ is the spread of the eigenfrequencies in the complex plane. Our results are not affected by the sign 
of the random variable $\epsilon_n^{(I)}$ thus unveiling a duality between gain ($\epsilon_n^{(I)}<0$) and lossy ($\epsilon_n^{(I)}>0$) structures. 

We point out that the effect of imaginary index of refraction on Anderson localization of light has been studied by a 
number of authors \cite{PMB96,RDVK00,JLS99}.  In all these cases, however, the authors were considering light localization along the propagation direction 
and their conclusions were based on the solutions obtained from the time-independent Schr\"odinger or Maxwell's equation. One of 
the main findings was that both gain and loss lead to the same degree of suppression of transmittance \cite{PMB96}. This counter-
intuitive duality was shown in Ref. \cite{JLS99}, using time-dependent Maxwell's equation, to be an artifact of time-independent calculations. 
Specifically it was shown that the amplitudes of both transmitted and reflected waves diverge due to lasing (in the case of gain) above 
a critical length scale. In contrast, in our set-up where localization is transverse to the paraxial propagation, divergence would not 
occur at any finite propagation distance and therefore the solutions of our problem are physically realizable.

{\it Physical set-up -- } We consider a one-dimensional array of weakly coupled single-mode optical waveguides. Light is transferred from waveguide to 
waveguide through optical tunneling. The propagation of light along the $z$-axis is described using coupled mode theory. The resulting equations are 
\cite{CLS03}
\begin{equation}
\label{eq2}
i\lambdabar{\partial \psi_n(z)\over \partial z}+V\left(\psi_{n+1}(z)+\psi_{n-1}(z)\right) + \epsilon_n \psi_n(z)=0
\end{equation}
where $n=1,\cdots,N$ is the waveguide number, $\psi_n(z)$ is the amplitude of the optical field envelope at distance $z$
in the $n$-th waveguide, $V$ is the tunneling constant between nearby waveguides (we assume below that $V=1$), $\lambdabar\equiv\lambda/2\pi$ where $\lambda$ 
is the optical wavelength, and $\epsilon_n=\epsilon_n^{(R)}+i\epsilon_n^{(I)}$ is the complex on-site effective index of refraction.  
Optical amplification can be introduced by stimulated emission in gain material or parametric conversion in nonlinear material, where as dissipation can be incorporated 
by depositing a thin film of absorbing material on top of the waveguide, or by introducing scattering loss in the waveguides. In order to 
distinguish the well understood Anderson localization phenomena which are associated with random $\epsilon_n^{(R)}$ from the localization 
phenomena related to the randomness of the imaginary part $\epsilon_n^{(I)}$, we consider below that all the 
waveguides have an identical effective index $\epsilon_n^{(R)}=\epsilon_0$ while $\epsilon_n^{(I)}$ is a random variable uniformly 
distributed in an interval $[-W;W]$. Due to the Kramers-Kronig relations the real and imaginary part of the dielectric constant 
are not independent of each other, nevertheless it is possible to have disorder only in the imaginary part by compensating for the 
changes in the $\epsilon_n^{(R)}$ by adjusting, for example, the width of the waveguides. The advantage offered by our system is 
the ability to study the dynamics of synthesized wavepackets, by launching an optical beam into any one waveguide or a 
superposition of any set of waveguides, and monitoring from the third dimension.

Substituting $\psi_n(z)=\phi_n \exp(-i\omega z)$, where $\omega$ can be complex, in Eq. (\ref{eq2})  we get the eigenvalue problem
\begin{equation}
\label{eq3}
\omega \phi_n =- (\phi_{n+1}+\phi_{n-1}) - \epsilon_n\phi_n
\end{equation}
In Fig. \ref{fig1} we report some typical FB modes for one realization of the disorder. We find that for sufficiently large disorder (or large system 
size) all modes are exponentially localized around some center of localization which can be either a gain (red) or a lossy (green) waveguide alike. 
The same qualitative picture applies also for the cases where all $\epsilon_n^{(I)}$ are positive (and random) or negative (and random). Therefore 
our set-up supports a {\it duality} between gain and loss. We want to quantify the structure of the FB modes of our system and identify the consequences of 
their transverse localization to the dynamics.

{\it  Exponential localization in the thermodynamic limit -- } We start our analysis by introducing the asymptotic participation number 
$\xi_{\infty}$ defined as
\begin{equation}
\label{asPN}
\xi_{\infty}(W,\omega) \equiv \lim_{N\rightarrow \infty}\langle\xi_N(W,\omega)\rangle
\end{equation}
Above the averaging has been performed over a number of disorder realizations and over FB modes inside a small frequency window around 
$\omega$. In all cases we had at least $8000$ data for statistical processing.

In Fig. \ref{fig2}(up) we report some representative data for the participation number $\langle\xi_N\rangle$, as a function of the system size 
$N$ for various disorder strengths $W$ and for $\omega=0$. One can extend the same analysis for other values of $\omega$ as well. From 
the data of Fig. \ref{fig2}(up) we have extracted the saturation value $\xi_{\infty}$. The results are summarized in Fig. \ref{fig2}(down) where we plot 
$\xi_{\infty}$ versus the disorder strength. Our analysis indicates that $\xi_{\infty}\sim 1/W^2$.  In case of exponentially localized 
FB modes, it is easy to show that, the asymptotic participation number Eq. (\ref{asPN}) is proportional to the inverse decay rate $\gamma$ of 
these modes.

\begin{figure}[h]
\includegraphics[scale=0.325]{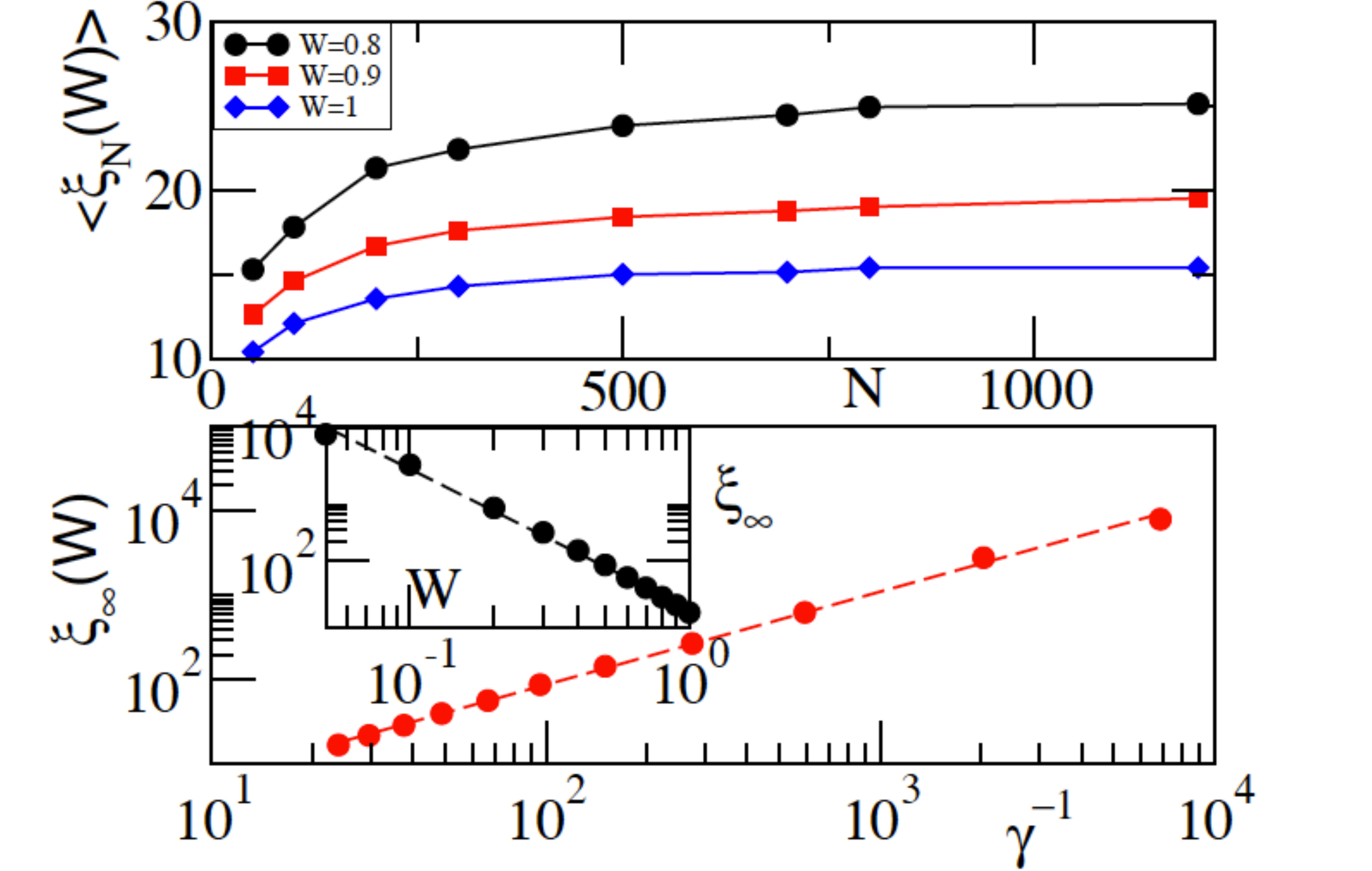}
\begin{center}
\caption{(Up) Scaling of the participation number $\langle\xi_N(W)\rangle$ vs. the system size $N$ for various disorder strengths $W$. A small energy
window around ${\cal R}e(\omega)=0$ such that ${\cal R}e(\omega)\in [-0.1, 0.1]$ has been considered. (Down) The extracted asymptotic participation
number versus the theoretical prediction of Eq. (\ref{lyapexact}) for the exponential decay rate $\gamma$. The best square fit (dashed line) gives
$\xi_{\infty}=0.55\gamma^{-1}$. In the inset we report the $\xi_{\infty}(W)$ versus the disorder strength $W$. The best square fit is $\xi_{\infty}(W)=
19.25 W^{-2}$.
\label{fig2}
}
\end{center}
\end{figure}

We shall now derive an explicit expression for the decay rate $\gamma(\omega)$ associated with a normal mode of frequency $\omega$. In 
order to obtain the transverse exponential growth (or decay) of the wavefunction amplitudes $\phi_n$ at sites $n$ we solve Eqs. (\ref{eq3}) 
recursively starting from some arbitrary value $\phi_1$, at site $n = 1$. We define:
\begin{equation}
\label{lyap}
\gamma \equiv -\lim_{n\rightarrow \infty} {1\over |n|}\langle \ln\left|{\phi_n\over \phi_1}\right|\rangle=-\lim_{N\rightarrow \infty} {1\over N}
\langle \sum_n^N\ln\left|R_n\right|\rangle
\end{equation}
where we have introduced the so-called Riccati variable $R_n\equiv {\phi_n\over\phi_{n-1}}$.  We can re-write Eq. (\ref{eq3}) as follows
\begin{equation}
\label{ricatti}
R_{n+1}+{1\over R_n} = (\omega-\epsilon_n)
\end{equation}
where now $\omega$ is considered an arbitrary frequency which we use as an input parameter \cite{note1}. Using Eqs. (\ref{lyap},\ref{ricatti}) 
we can then evaluate numerically $\gamma(\omega)$.

Next we write $R_n$ as $A\times \exp(W B_{n}+W^{2} C_n+...)$ and substitute in Eq. (\ref{ricatti}) $\omega=2 \cos q$, where $q$ is in general 
a complex quantity. For weak disorder we can further expand $R_n$ in Taylor series of $W$. Equating the same powers of $W$ in Eq. (\ref{ricatti})
while taking into consideration the statistical nature of $\epsilon_{n}$ (e.g. $\langle\epsilon_{n}\rangle=0$), we get expressions for $A$, $\langle 
B_n\rangle$, $\langle B_n^2\rangle$ and $\langle C_n\rangle$ as a function of $W$. Substituting them back to Eqs. (\ref{lyap},\ref{ricatti}) we get,
up to second order in $W$, that
\begin{equation}
\label{lyapexact}
\gamma= q_I + (\frac{W^{2}}{24}) {\frac{\omega_I^{2}\coth^{2}(q_I)-
 \omega_R^{2}\tanh^{2}(q_I)}{(\frac{1}{4}) [\omega_I^{2}\coth^{2}(q_I)-
 \omega_R^{2}\tanh^{2}(q_I)]^{2}+\omega_I^{2}\omega_R^{2}}} 
\end{equation}
where $\omega_R={\cal R}e(\omega); \omega_I={\cal I}m(\omega); q_R={\cal R}e(q); q_I={\cal I}m(q)$. A comparison
between the theoretical expression Eq. (\ref{lyapexact}) and the numerically extracted asymptotic participation number $\xi_{\infty}$ is shown
in Fig. \ref{fig2}(down).

{\it One Parameter Scaling Theory --}  We are now ready to formulate a one-parameter scaling theory of the finite length participation number 
of the FB modes of our system Eq. (\ref{eq3}). To this end we postulate the existence of a function $f(\Lambda)$ such that
\begin{equation}
\label{scale1}
p_N(W) = f(\Lambda)\quad {\rm where} \quad \Lambda\equiv {\xi_{\infty}\over N}
\end{equation}
where $p_N(W)$ is defined in Eq. (\ref{eq1}). It is then straightforward to show 
that Eq. (\ref{scale1}) can be written equivalently in the form of Eq. (\ref{eq1}) \cite{fnote}.
In the localized regime $\Lambda\ll 1$ (infinite system sizes $N$) the finite length participation number $\xi_N(W,\omega)$ has to converge 
to its asymptotic value $\xi_{\infty}(W,\omega)$, see Eq. (\ref{asPN})); thus we expect that $f(\Lambda)\rightarrow \Lambda$. In the other 
limiting case $\Lambda\gg 1$, corresponding to the de-localized regime, we have that $\xi_N(W)\propto 2N/3$ (i.e. the wave-functions extend 
over the whole available space) and thus $f(\Lambda) \rightarrow 1$ \cite{note3}. 

We have confirmed numerically the validity of Eq. (\ref{scale1}) for our system Eq. (\ref{eq3}). The numerical data are shown in Fig. \ref{fig3}. 
Various values of $N$ in the range $100-1200$ have been used while the width of the box distribution of the random imaginary refraction 
indexes $W$ was taken in the range $0.05\leq W\leq 1$. We have also checked (not shown here) that the same
scaling behavior is applicable for the case where $n_I$ takes random values which are only positive or negative.

\begin{figure}[h]
\includegraphics[scale=0.325]{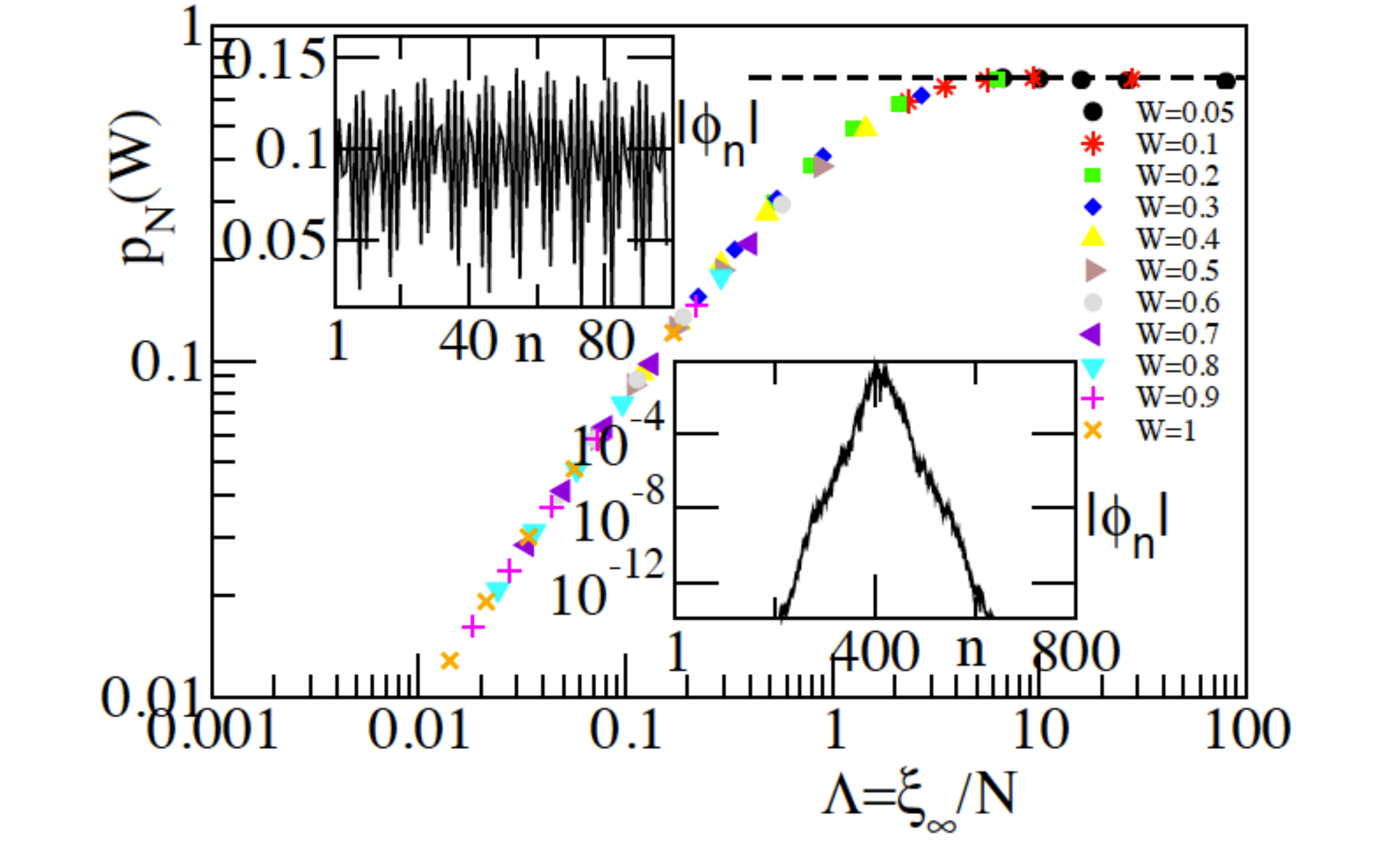}
\begin{center}
\caption{Scaled participation ratio $p_N(W)\equiv \xi_N/N$ vs. the scaling parameter $\Lambda \equiv\xi_{\infty}/N$ for various $N$-values and disorder
strengths $W=0.1-1$. The eigenmodes are taken from a small frequency window at the center of the band. Insets: Two typical FB modes in the localized
(lower left) and in the delocalized (upper right) domain. The dashed line is the theoretical value of $2/3$ for the limiting case of $\Lambda\gg1$.
\label{fig3}}
\end{center}
\end{figure}

{\it Autocorrelation function and Break-length --} A natural question that can be raised is how the transverse localization of the Floquet-Bloch modes 
of the coupled waveguide array of Eq. (\ref{eq2}) is reflected in the paraxial propagation of a beam which is initially localized at some waveguide 
$n_0$. A dynamical observable that can be used in order to trace the effects of localization is the return to the origin probability $P_{n_0}(z)=
|\psi_{n_0}(z)|^2\equiv \langle n_0|\psi(z)\rangle|^2$. In the case of lossless random media this quantity has been used in order to quantify the 
degree of Anderson localization. Specifically it can be shown that in this case $P_{n_0}(z\rightarrow \infty)\rightarrow \xi_{\infty}^{-1}$. In contrast, 
in the case of periodic lattices $P_{n_0}(z)=|J_0(2Vz)|^2$ where $J_0(x)$ is the zero-th order Bessel function. Since $P_{n_0}(z)$ is a fluctuating 
quantity, we often investigate its smoothed version $C(z)=(1/z)\int_0^z P(z') dz'$. For periodic lattices $C(z)\sim 1/z$, indicating a 
loss of correlations of the evolving beam with the initial preparation.

We have introduced a rescaled version of $C(z)$ such that it takes into account the growth/loss of the total field intensity due to the presence of 
the dissipative part of the index of refraction at the waveguides
\begin{equation}
\label{Crescale}
C(z)={1\over z}\int_0^z P(z')dz'/I(z),\quad{\rm where}\quad I(z)=\sum_n|\psi_n(z)|^2
\end{equation} 
and compare its deviations from the ballistic results $C_{\rm bal}(z)\sim1/z$ corresponding to a perfect lattice \cite{note2}. We have found that the 
correlation function of the disordered lattice follows the ballistic results up to a propagation distance $z^*$ which depends on the disorder $W$ of 
$n^{(I)}$. We determined the break-length $z^*$ by the condition $Q(z)=(C(z)/C_{\rm bal}(z))-1=0.1$ which correspond to $10\%$ deviations of $C(z)$ 
from the behavior shown by the perfect lattice. To suppress the ensemble fluctuations further, we averaged $C(z)$ over more than $50$ different disorder 
realizations. Then the (averaged) break-length $z^*$ is determined by the condition $\langle Q(z^*)\rangle=0.1$. The dependence of $\langle Q(z)
\rangle$ on distance, for representative disorder widths $W$, is shown in Fig. \ref{fig4}(up). We find that $z^*$ becomes smaller as we increase the disorder 
$W$. The numerically extracted $z^*$ values and their dependence on $W$ is summarized in Fig. \ref{fig4}(down).  The fit of the numerical data gives a 
power law dependence $z^*\approx W^{-\alpha}$ with $\alpha\approx 1.35\pm 0.02$, being quite robust to other definitions (e.g. $5\%$ deviation level) 
of break length.

\begin{figure}[h]
\includegraphics[scale=0.325]{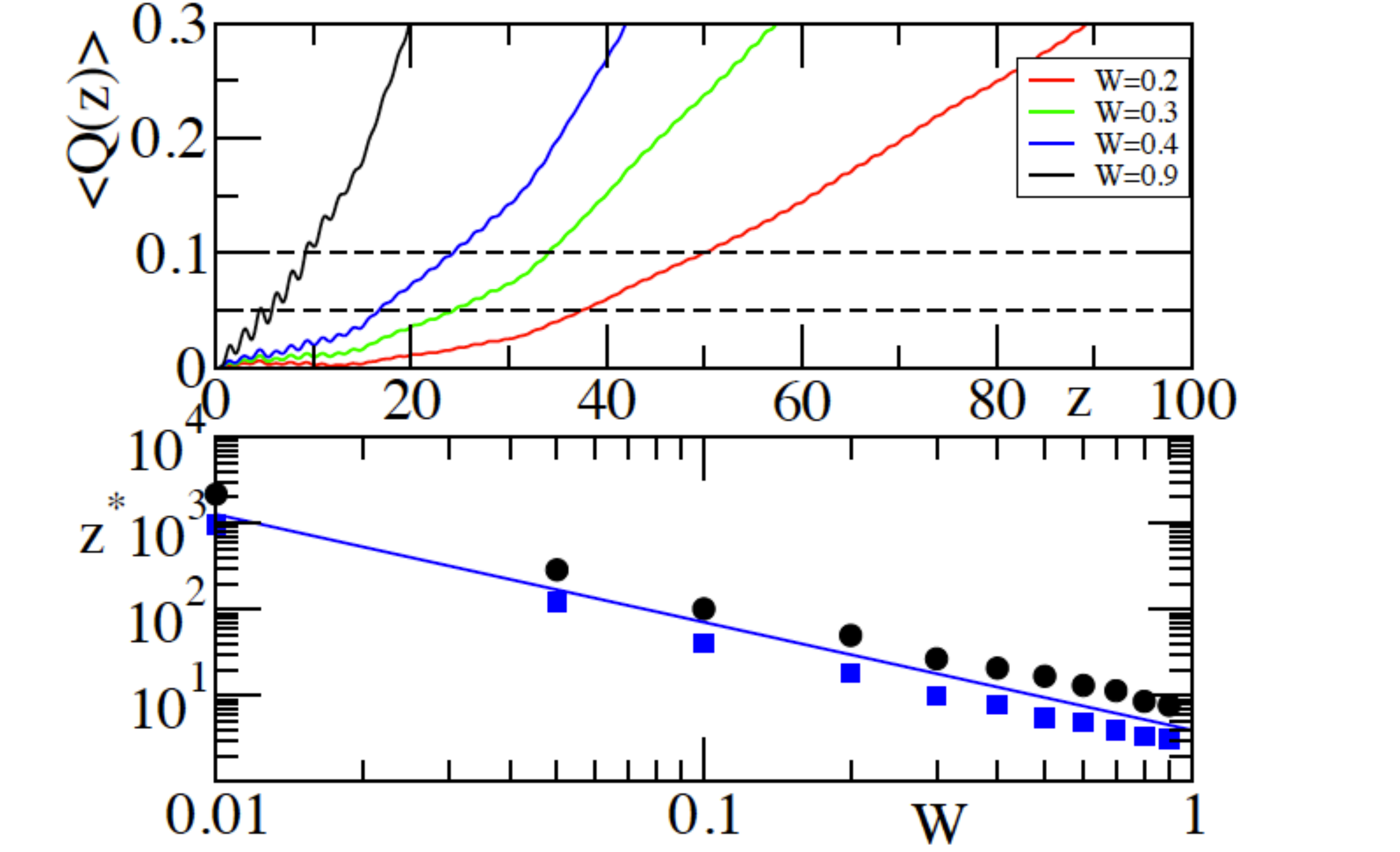}
\begin{center}
\caption{(Up) The averaged $\langle Q(z)\rangle$ function versus distance $z$ for some typical values of disorder strength $W$. The horizontal black 
dashed lines indicate the $5\%$ and $10\%$ deviations of $C(z)$ from the ballistic result $C_{\rm bal}(z)$. (Down) The break-length $z^*$ versus the 
disorder strength $W$ for $5\%$ (blue squares) and $10\%$ (black circles) deviations. The straight line is the best fit and has a slope $-1.35$.
\label{fig4}}
\end{center}
\end{figure}

The following argument provides some understanding of the dependence of the break-length on the disorder strength. Our explanation is
based on the fact that in a non-Hermitian system the physics is affected by the distribution of the complex frequencies of the effective non-Hermitian
Hamiltonian that describes the paraxial evolution of the beam in the waveguide array. 

Once the disorder $W$ is introduced to the imaginary part of the refraction indexes, the eigenfrequencies acquire an imaginary part that determines 
the growth/decay of the associated normal modes of the system.  We consider - to a good approximation - that they are distributed homogeneously 
in a narrow strip around the real axis with area ${\cal A}\sim \Delta {\cal R}e(\omega) \cdot \Delta {\cal I}e(\omega)$. The length of the box is fixed 
$\Delta {\cal R}e(\omega)\propto 2V$ while its width $\Delta {\cal I}m(\omega)$ depends on the disorder strength as $\Delta {\cal I}e(\omega)
\propto W$. Therefore we have that ${\cal A}\sim W$.

Since, on the other hand, the FB modes are localized then only $\xi_{\infty}$ out of them have a significant overlap with the initial localized state 
and thus effectively participate in the evolution. Their effective frequency spacing in the complex plane $\delta$ defines the energy scale that 
determines the deviations from the periodic lattice behavior. The associated break-length is defined as $z^*\sim 1/\delta$. The latter is estimated 
by realizing that $\xi_{\infty} \delta^2 \approx {\cal A}$. Solving with respect to $\delta$ we get
\begin{equation}
\label{delta}
\delta \sim \sqrt{{\cal A}/\xi_{\infty}}\sim W^{1.5}\rightarrow z^*\sim 1/\delta\sim W^{-1.5}
\end{equation}
Above we have substituted ${\cal A}\approx \Delta {\cal R}e(\omega) \cdot \Delta {\cal I}m(\omega)\sim W$ and use that $\xi_{\infty}\sim W^{-2}$. 
The theoretical dependence Eq. (\ref{delta}) is slightly different but very close to the numerical value $1.35$ that we got from the best square fit of the 
data of Fig. \ref{fig4}(down). We attribute this difference to the fact that the localization length that we have used in Eq. (\ref{delta}) is the one associated 
with the modes around ${\cal R}e(\omega)\approx 0$ while for other frequencies might scale as $\xi_{\infty}\sim 1/W^{\mu}$ with $\mu<2$ 
(Wegner-Kappus resonances). Since, however, an initial $\delta$-like beam will excite FB modes with various frequencies it is more appropriate to 
introduce an average localization length over all  frequencies and consider the scaling of $\Delta {\cal I}m(\omega)$ over the whole spectrum. A 
scaling analysis along these lines (see supplementary material) indicates that $\xi_{\infty}\sim 1/W^{1.27}$ and $\Delta {\cal I}m(\omega)\sim W^{1.55}$ 
which leads to $z^*\sim W^{-1.4}$. 

\textit {Conclusions -} In conclusion we have demonstrated that randomness only in the dissipative part of the impedance profile of a medium can 
result in localization. Using an array of coupled waveguides as a prototype for this class of systems, we have established a renormalization approach 
for the localization properties of the FB modes of the effective non-Hermitian Hamiltonian that describes the paraxial evolution of light in the array
and  illustrated their consequences in the light propagation.

\textit{Acknowledgement - } This work was sponsored partly by an NSF ECCS-1128571 grant and by an AFOSR MURI grant FA9550-14-1-0037

\clearpage
\section{Supplemental Material}

\section{Derivation of Eq. (\ref{eq1}) from Eq. (8)}
Taking the derivative of Eq. (8) with respect to $\ln(N)$ 
we get that $\partial p_N(W)/\partial\ln N = -\Lambda \partial f(\Lambda)/\partial \Lambda=F(\Lambda)$. Substituting $\Lambda=f^{-1}(p_N(W))$ 
back to the latter equation allows us to rewrite the right hand side of it as $F(\Lambda=  f^{-1}(p_N(W))) = \beta(p_N(W))$ which proves the validity 
of Eq. (\ref{eq1}).

\vspace*{1cm}
\section{Scaling quantities after averaging over the whole spectrum}
In this section we have investigated the scaling of localization length $\xi_{\infty}$ versus the disordered strength $W$ when the averaging over the 
eigenmodes of the effective Hamiltonian Eq. (3) is performed over the whole frequency spectrum. Our starting point is the definition in Eq. (4): 
\begin{equation}
\label{S1}
\tag{$S1$}
{\bar \xi_{\infty}}(W)=\lim_{N\rightarrow \infty} \langle{\bar \xi_N}(W)\rangle
\end{equation}
where $\langle\cdots\rangle$ indicates the standard averaging over disorder realizations and ${\bar {\cdots}}$ the additional averaging over the 
whole freqeuncy spectrum. Some representative data for the finite participation number ${\bar \xi_N}(W)$ versus the system size $N$ are shown 
in Fig. 5(up). A summary of the extracted asymptotic values ${\bar \xi_{\infty}}(W)$ are shown in Fig. 5(down). The best square fit indicates 
that  
\begin{equation}
\label{S2}
\tag{$S2$}
{\bar \xi_{\infty}}(W) \sim W^{-1.27}
\end{equation}

\begin{figure}[tbp]
\includegraphics[scale=0.35]{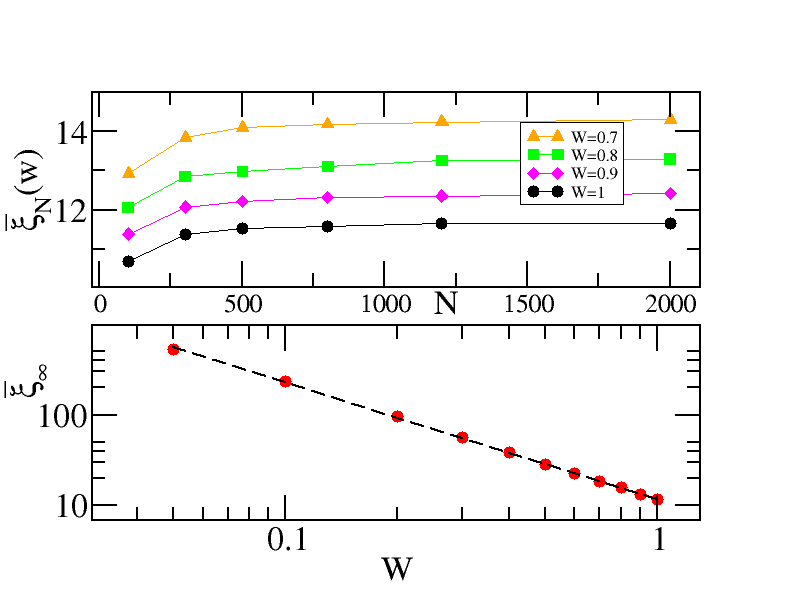}
\renewcommand{\figurename}{Fig. }
\caption{(Up) Asymptotic behavior of the participation number $\langle\xi_N(W)\rangle$ for the large system sizes and various disorder strengths $W$. 
These data cover the whole energy window where ${\cal R}e(\omega)\in [-2, 2]$. (Down) Asymptotic participation number $\xi_{\infty}(W)$ versus the 
disorder strength $W$ follows a scaling as $\xi_{\infty}(W)\sim W^{-\mu}$ with $\mu=1.27\pm0.02$ given by the best square fit.
\label{figS1}
}
\end{figure}

We have also confirmed the validity of Eq. (\ref{S2}) by establishing that it is the appropriate variable for the applicability of the one-parameter scaling 
theory of the participation number in the case where the averaging is performed over the whole spectrum. The associated rescaled participation number 
${\bar p}_N(W)\equiv {\bar \xi_N}/N$ versus the scaling parameter ${\bar \Lambda}\equiv {\bar \xi}_{\infty}(W)/N$ is reported in 
Fig. 6. A nice 
scaling is evident.

\begin{figure}[tbp]
\includegraphics[scale=0.35]{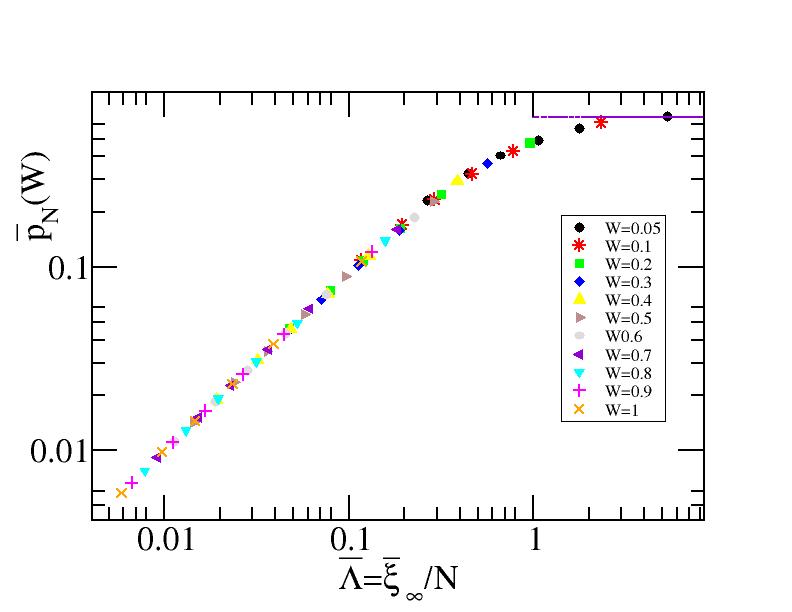}
\renewcommand{\figurename}{Fig. }
\caption{Scaled participation ratio $p_N(W)\equiv \xi_N/N$ vs. the scaling parameter $\Lambda \equiv\xi/N$ for various $N$'s and disorder
strengths $W=0.05-1$. The eigenmodes belong to whole frequency range. The theoretical value of $2/3$ (maroon line) is confirmed 
for the limiting case of $\Lambda\gg1$ .
\label{figs2}}
\end{figure}

Armed with the above knowledge of Eq. (\ref{S2}) we apply the argument of Eq. (10) and re-evaluate the prediction of break-time $z^*$,
under the (more realistic) assumption that all modes participate in the evolution of the wavepacket. To this end we first evaluate numerically the 
variance $\sigma^2_{{\cal I}m(\omega)}\sim (\Delta {\cal I}(\omega))^2$ of the imaginary part of the complex frequencies of the non-Hermitian Hamiltonian 
(3). In Fig. 7(up) we depicted a typical distribution of eigenvalues of our Non-Hermitian Hamiltonian. Notice that modes at the edges of the 
band move further up/down in the complex plain, thus invalidating the assumption of a uniform distribution of $\omega_n$'s in the complex plane 
around the real axis (this assumption is still valid as long as we concentrate on a small freqeuncy window around ${\cal R}e(\omega)=0$). In Fig. 7(down),
 the scaling of the standard deviation $\sigma_{{\cal I}m(\omega)}\sim \Delta{\cal I}m(\omega)$ is presented vs. the disorder amplitude $W$. We find the following
scaling relation
\begin{equation}
\label{S3}
\tag{$S3$}
\sigma_{{\cal I}m(\omega)}\sim W^{1.55}
\end{equation}

Finally substituting Eqs. (\ref{S2},\ref{S3}) in Eq. (10) we find that $z^*\sim W^{-1.4}$ in nice agreement with the results of the numerical analysis of Fig.4
of the main text.

\begin{figure}[tbp]
\includegraphics[scale=0.35]{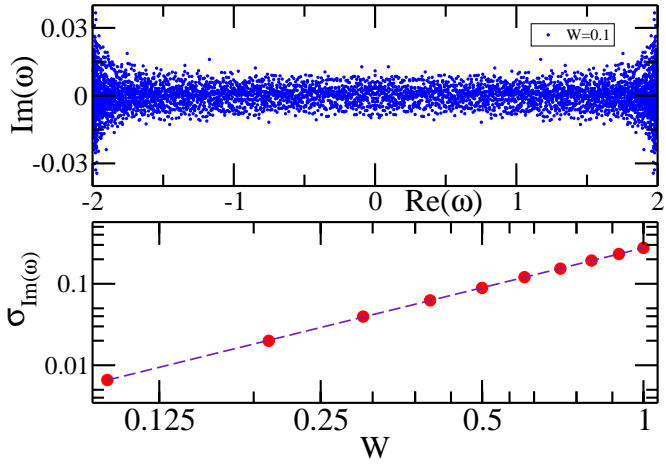}
\begin{center}
\renewcommand{\figurename}{Fig. }
\caption{(Up) Imaginary vs. real parts of eigenvalues for N=500. (Down) standard deviation of ${{\cal I}m(\omega)} $ as a function of W. The best square fit
indicates $\sigma_{{\cal I}m(\omega)}\sim W^{1.55}$
\label{figs3}}
\end{center}
\end{figure}


\begin{thebibliography}{99}
\bibitem{note1} The reasoning for investigating $\phi_n$ at any frequency $\omega$ rather than at the eigenfrequencies 
is based on the Borland conjecture \cite{B63} which states that when $\omega$ is close to an eigenfrequency the exponential 
growth/decay of $\phi_n$ at distant sites converges to the exponential rate of amplitude variation of the corresponding 
eigenstate around its localization center.
\bibitem{note3} In the delocalized limit the wavefunctions of a perfect lattice are $\phi_n^{(k)}=\sqrt{2\over N+1} \sin({nk\pi\over N+1})$
where $n,k=1,\cdots,N$. Then the participation number, in the large $N$ limit, can be evaluated analytically to be $\xi_N\approx {2\over 3}N$.
\bibitem{note2} The power law behaviour $C(z)\sim 1/z$ continues to hold also in the case of uniformly lossy or gain periodic 
waveguide arrays once the rescale with the temporal norm Eq. (\ref{Crescale}) is taken into consideration.
\end{thebibliography}
\end{document}